\documentclass[epj,nopacs]{svjour}
\usepackage{graphics}
\begin{document}
\title{On jet structure in heavy ion collisions}
\author{I.P.~Lokhtin, A.A.~Alkin, A.M.~Snigirev}  
%
%
\institute {Skobeltsyn Institute of Nuclear Physics, Lomonosov Moscow State 
University, Moscow, Russia }
%
%

\abstract{The LHC data on jet fragmentation function and jet shapes in PbPb collisions at
center-of-mass energy 2.76 TeV per nucleon pair are analyzed and interpreted in the 
frameworks of PYQUEN jet quenching model. A specific modification of 
longitudinal and radial jet profiles in most central PbPb collisions as compared with pp data 
is close to that obtained with PYQUEN simulations taking into account wide-angle radiative and 
collisional partonic energy loss. The contribution of radiative and collisional loss to the 
medium-modified intra-jet structure is estimated. 
%
}
%
\titlerunning{On jet structure in heavy ion collisions}
\authorrunning{I.P.~Lokhtin, A.A.~Alkin, A.M.~Snigirev}
\maketitle

\section{Introduction}
\label{sec:intro}

Studying the modification of jets as they are formed from energetic partons
propagating through the hot and dense medium created in ultrarelativistic heavy 
ion collisions is a particularly useful tool for probing the produced matter's 
properties. The energy loss of jet partons in deconfined medium, quark-gluon plasma, 
is predicted much 
stronger than in cold nuclear matter, and leads to so called ``jet quenching'' effect
(see, e.g., reviews~\cite{d'Enterria:2009am,Wiedemann:2009sh,Accardi:2009qv,Majumder:2010qh,Dremin:2010jx,Mehtar-Tani:2013pia,Majumder:2014vpa} and references therein).  
Indirectly jet quenching was observed for the first time at RHIC experiments via 
measurements of inclusive high-$p_T$ hadron production in gold-gold collisions at  
center-of-mass energy $\sqrt s_{\rm NN}=200$ GeV. It was manifested as the suppression of overall 
high-$p_T$ hadron rates and back-to-back dihadron azimuthal correlations (including the specific 
azimuthal angle dependence of the effect with respect to the event plane). The summary 
of experimental results at RHIC can be found in~\cite{brahms,phobos,star,phenix}. 

The lead-lead collision energy at LHC, $\sqrt s_{\rm NN}=2.76$ TeV, is a factor of $\sim 14$ 
larger than that in RHIC, thereby allows one to probe new frontiers of super-hot quark-gluon
matter. The analysis of high statistics samples of fully reconstructed jets becomes possible.
The results of jet analyses of 2010 and 2011 PbPb data obtained by three LHC experiments, 
ALICE, ATLAS and CMS, is summarized e.g. in review~\cite{Spousta:2013aaa}.
The first direct evidence of jet quenching has been observed as the large, 
centrality-dependent, imbalance in dijet transverse 
energy~\cite{Aad:2010bu,Chatrchyan:2011sx}. It has been found using missing 
$p_{\rm T}$ techniques that the jet energy loss spreads over low transverse momenta and 
large angles~\cite{Chatrchyan:2011sx}. Then jet momentum dependence of the dijet imbalance 
has been studied in detail~\cite{Chatrchyan:2012nia}. Significant 
transverse energy imbalance has been observed for photon+jet production in PbPb 
events~\cite{Chatrchyan:2012gt}. Another direct manifestation of jet quenching is the 
inclusive jet suppression in central PbPb collisions compared to peripheral 
events~\cite{Aad:2012vca,Abelev:2013kqa} and to proton-proton interactions~\cite{Aad:2014bxa}. 
The azimuthal angle dependence of such suppression has also been measured~\cite{Aad:2013sla}. 
The similar level of suppression as for inclusive jets is seen for jets from b-quark 
fragmentation~\cite{Chatrchyan:2013exa}. In contrast to PbPb collisions, proton-lead data from LHC 
do not show jet quenching manifestation~\cite{Chatrchyan:2014hqa}. 

A number of theoretical calculations and Monte-Carlo simulations were attempted to reproduce 
some of basic features of jet quenching pattern at LHC (dijet and 
photon+jet imbalance, nuclear modification factors for jets and high-$p_{\rm T}$ 
hadrons), and to extract by such a way the information about medium properties and partonic energy loss mechanisms~\cite{CasalderreySolana:2010eh,Qin:2010mn,Young:2011qx,Srivastava:2011nq,Lokhtin:2011qq,Lokhtin:2012re,Betz:2012qq,Renk:2012cx,Renk:2012cb,Renk:2013rla,Apolinario:2012cg,Zapp:2012ak,Dai:2012am,Huang:2013vaa,Zakharov:2012fp,Zakharov:2013gya,Burke:2013yra,Xu:2014ica,Mehtar-Tani:2014yea,Casalderrey-Solana:2014bpa}. 
The observable that allows more precise comparison between the data and theoretical models of jet quenching to be done, is the internal jet structure. The recently published experimental
data on medium-modified jet structure include the measurement of jet shapes (radial 
profile)~\cite{Chatrchyan:2013kwa} and jet fragmentation function (longitudinal 
profile)~\cite{Chatrchyan:2014ava,Aad:2014wha}. It was suggested in~\cite{Cacciari:2012mu} 
that it could be also of interest to study moments of jet fragmentation function. 

In present paper, the LHC data on jet fragmentation function and jet shapes in PbPb 
collisions are analyzed and interpreted in the frameworks of PYQUEN partonic energy loss 
model~\cite{Lokhtin:2005px}. In the previous papers~\cite{Lokhtin:2011qq,Lokhtin:2012re} 
this model were applied to analyze the dijet energy asymmetry and nuclear modification 
factor of inclusive hadrons.   

\section{PYQUEN model}
\label{sec:model}

PYQUEN (PYthia QUENched) is one of the first Monte-Carlo models
of jet quenching~\cite{Lokhtin:2005px}. PYQUEN was constructed as a modification 
of jet events obtained with the generator of hadron-hadron interactions 
PYTHIA$\_$6.4~\cite{Sjostrand:2006za}. The details of the used physics model 
and simulation procedure can be found in the original paper~\cite{Lokhtin:2005px}. 
Main features of the model are listed below as follows.

The approach describing the multiple scattering of hard partons relies on 
accumulated energy loss via gluon radiation, which is associated with each
parton scattering in a hot matter. It also includes
the interference effect in gluon emission with a finite formation
time using the modified radiation spectrum $dE/dl$ as a function
of the decreasing temperature $T$.  The basic kinetic integral equation for 
the partonic energy loss $\Delta E$ as a function of initial energy $E$ and path 
length $L$ has the form 
\begin{eqnarray} 
\label{elos_kin}
\Delta E (L,E) = \int\limits_0^Ldl\frac{dP(l)}{dl}
\lambda(l)\frac{dE(l,E)}{dl} \, , \\ 
\frac{dP(l)}{dl} = \frac{1}{\lambda(l)}\exp{\left( -l/\lambda(l)\right) }
\, ,  \nonumber
\end{eqnarray} 
where $l$ is the current transverse coordinate of a parton, $dP/dl$ is the 
scattering probability density, $dE/dl$ is the energy loss per unit length, 
$\lambda = 1/(\sigma \rho)$ is the in-medium mean free path, $\rho \propto T^3$ is 
the medium density at the temperature $T$, $\sigma$ is the integral cross 
section for the parton interaction in the medium. 

The model takes into account radiative and collisional energy loss of hard partons in
longitudinally expanding quark-gluon fluid, as well as the
realistic nuclear geometry. The radiative energy loss is treated in the 
frameworks of BDMPS model~\cite{Baier:1996kr,Baier:1999ds,Baier:2001qw}: 
\begin{eqnarray} 
\label{radiat} 
\frac{dE}{dl}^{rad} = \frac{2 \alpha_s (\mu_D^2)C_R}{\pi L}
\int\limits_{\mu_D^2\lambda_g}^E  
d \omega \left[ 1 - y + \frac{y^2}{2} \right] 
\>\ln{\left| \cos{(\omega_1\tau_1)} \right|} 
\>, \\  
\omega_1 = \sqrt{i \left( 1 - y + \frac{C_R}{3}y^2 \right)   
\bar{\kappa}\ln{\frac{16}{\bar{\kappa}}}}~,~~ 
\bar{\kappa} = \frac{\mu_D^2\lambda_g  }{\omega(1-y)} ~, \nonumber
\end{eqnarray} 
where $\tau_1=L/(2\lambda_g)$, $y=\omega/E$ is the fraction of the hard parton 
energy carried away by the radiated gluon, $\alpha_s$ is the QCD running 
coupling constant for $N_f$ active quark flavors, $C_R = 4/3$ is the quark color 
factor, and $\mu_D$ is the Debye screening mass. A similar expression for the 
gluon jet can be obtained by setting  $C_R=3$ and properly changing the factor 
in the square brackets in 
(\ref{radiat})~\cite{Baier:1999ds}. The simple generalization to a radiative 
energy loss of massive quark case uses the ``dead-cone'' 
approximation~\cite{Dokshitzer:2001zm}. 

The collisional energy loss due to elastic scatterings is calculated in the 
high-momentum transfer limit~\cite{Bjorken:1982tu,Braaten:1991jj,Lokhtin:2000wm}: 
\begin{eqnarray}
\label{col} 
\frac{dE}{dl}^{col} = \frac{1}{4T \lambda \sigma} 
\int\limits_{\displaystyle \mu^2_D}^
{\displaystyle t_{\rm max}}dt\frac{d\sigma }{dt}t \, ,
\end{eqnarray} 
where the dominant contribution to the differential scattering cross section is 
\begin{eqnarray} 
\label{sigt} 
\frac{d\sigma }{dt} \cong C \frac{2\pi\alpha_s^2(t)}{t^2} 
\frac{E^2}{E^2-m_p^2} \, 
\end{eqnarray} 
for the scattering of a hard parton with energy $E$ and mass $m_p$ off the ``thermal'' 
parton with energy (or effective mass) $m_0 \sim 3T \ll E$. Here $C = 9/4, 1, 4/9$ 
for $gg$, $gq$ and $qq$ scatterings respectively. The integrated cross section $\sigma$ 
is regularized by the Debye screening mass squared $\mu_D^2 (T) \simeq 4\pi 
\alpha _s T^2(1+N_f/6)$. The maximum momentum transfer is 
$t_{\rm max}=[ s-(m_p+m_0)^2] [ s-(m_p-m_0)^2 ] / s$ where $s=2m_0E+m_0^2+m_p^2$. 

The medium where partonic rescattering occurs is considered as a boost-invariant 
longitudinally expanding perfect quark-gluon fluid, and the partons as being produced on a 
hyper-surface of equal proper times $\tau$~~\cite{bjork86}. Then the proper time 
dependence of a temperature $T$, energy density $\varepsilon$  and number density 
$\rho$ gets the form:
\begin{equation}
\varepsilon(\tau) = \varepsilon_0 \left( \frac {\tau_0}{\tau} \right) ^{4/3},~
T(\tau)  = T_0 \left( \frac{\tau_0}{\tau}\right) ^{1/3},~ \rho(\tau)= \rho_0 
\frac{\tau_0}{\tau}~.
\end{equation}
In principle, other scenarios of space-time evolution of quark-gluon matter can be considered 
within PYQUEN model~\cite{Lokhtin:2005px}. For example, the presence of fluid viscosity slows 
down the cooling rate, i.e., in fact the effective temperature gets higher as compared with the 
perfect fluid case. At the same time, the influence of the transverse expansion of the medium 
on parton rescattering intensity is practically inessential for high initial temperatures 
$T_0^{\rm max}$. 

The strength of partonic energy loss in PYQUEN is determined mainly by the initial maximal 
temperature $T_0^{\rm max}$ of hot fireball in central PbPb collisions, which is achieved 
in the center of nuclear overlapping area at mid-rapidity. The transverse energy density in 
each point inside the nuclear overlapping zone is supposed to be proportional to the 
impact-parameter dependent product of two nuclear thickness functions $T_A$ in this point: 
\begin{eqnarray}
\label{enprof} 
& & \varepsilon (r_1, r_2) \propto T_A(r_1) T_A(r_2)~,  
\end{eqnarray}
where $r_{1,2}$ are the transverse distances between the centres of colliding nuclei and the 
parton production vertex. The rapidity dependent spreading of the initial energy density around 
mid-rapidity $y=0$ is taken in the Gaussian-like form. The partonic energy loss depends also on 
the proper time $\tau_0$ of matter formation and the number $N_f$ of active flavors in the 
medium. Note that the variation of $\tau_0$ value within its reasonable range has rather moderate 
influence on the strength of partonic energy loss. The jet quenching gets stronger at larger 
$\tau_0$ due to slower medium cooling, which implies the jet partons spending more time in the 
hottest regions, and as a result the rescattering intensity some increases~\cite{Lokhtin:2011qq}. 

Another important ingredient of the model is the angular spectrum of medium-induced 
radiation. There is a number of recent theoretical achievements related to this subject 
(see, e.g., \cite{Zapp:2012ak,CasalderreySolana:2012ef,Ramos:2014mba,Kurkela:2014tla,Blaizot:2014ula}). 
Nevertheless to the best of our knowledge there is no unique well-defined angular spectrum 
of emitted gluons in the current literature to be implemented in Monte-Carlo models 
unambiguously. This is one of main reasons leading to various approaches to modeling of 
medium-induced partonic energy loss in the existing event generators. Since the detailed 
Monte-Carlo treatment of the angular spectrum of medium-induced radiation is rather 
sophisticated and ambiguously determined, the simple parameterizations of the gluon distribution 
over the emission angle $\theta$ are used within PYQUEN, to get some notion about possible 
effects related to angular structure and to illustrate our interpretation of the data. 
There are two basic limiting options in the model. The first one is the ``small-angle'' radiation,
\begin{equation} 
\label{sar} 
\frac{dN^g}{d\theta}\propto \sin{\theta} \exp{\left( 
-\frac{(\theta-\theta_0)^2}{2\theta_0^2}\right) }~, 
\end{equation} 
where $\theta_0 \sim 5^0$ is the typical angle of the coherent gluon radiation as estimated 
in~\cite{Lokhtin:1998ya}. This scenario results in very close predictions to those obtained under 
assumption that all gluons are emitted collinearly (along the direction of motion of radiating 
particle). The second option is the ``wide-angle'' radiation,  
\begin{equation} 
\label{war} 
\frac{dN^g}{d\theta}\propto 1/\theta~, 
\end{equation} 
which is similar to the angular spectrum of parton showering in a vacuum without coherent 
effects~\cite{Dokshitzer:1991wu}. The physical meaning of wide-angle radiation could 
be the presence of intensive secondary rescatterings of in-medium emitted 
gluons~\cite{Qin:2010mn}. It may result in destroying the small-angle (BDMPS-like)
coherence emission and significant broadening of gluon emission angular spectrum. This scenario 
allows us to estimate effects of radiation outside the typical jet cone.
The collisional energy loss in PYQUEN is always ``out-of-cone'' loss. 
Such lost energy is considered as ``absorbed'' by the medium, because the major part of  
``thermal'' particles knocked out of the hot matter by elastic rescatterings fly 
outside the typical jet cone~\cite{Lokhtin:1998ya}. 

The event-by-event Monte-Carlo simulation procedure in PYQUEN includes three main steps: 
the generation of initial parton spectra with PYTHIA and production vertexes at 
the given impact parameter; the rescattering-by-rescattering passage of each jet 
parton through a dense zone accompanied with gluon radiation and collisional energy 
loss; the final hadronization for jet partons and in-medium emitted gluons according 
to the standard Lund string scheme implemented in PYTHIA. 
 
\section{Results}
\label{sec:results}

PYQUEN model was applied to simulate medium-modified inclusive jet production at 
energy $\sqrt s_{\rm NN}=2.76$ TeV for different PbPb centralities. The radiative and 
collisional energy loss were taken into account. Two options for the angular spectrum of 
gluon radiation were considered, wide- and small-angle radiative loss. Hereafter let 
us call these options as ``Scenario {\it W}'' and ``Scenario {\it S}'' respectively. 
The discrimination between these two ultimate scenarios when analyzing various jet characteristics 
seems particularly interesting, because it allows one to get some notion about possible effects 
related to 
angular structure of medium-induced partonic energy loss. ``Scenario {\it S}'' results in very 
close predictions to those obtained under assumption that gluons are emitted collinearly, while
 ``Scenario {\it W}'' is useful to estimate effects of radiation outside the typical jet cone.
At the same time a specific form of angular spectrum is not crucial for these two extreme cases, 
only a part of the energy loss outside the typical jet cone is significant for the effects  
under consideration. 

PYQUEN parameter values $T_0^{\rm max}=1$ GeV, $\tau_0=0.1$ fm$/c$ and $N_f=0$ (gluon-dominated 
plasma) were used for our simulations. Such parameter settings allow the model to 
reproduce the LHC data on dijet transverse energy imbalance and nuclear modification 
factors of inclusive hadrons as wide-angle radiative and collisional energy loss are 
taken into account~\cite{Lokhtin:2011qq,Lokhtin:2012re}. At that PYTHIA tune Pro-Q20 was utilized. 

In the current paper we do not intend to do a complete ``apples to apples'' comparison of the model 
results with the data. Such kind of comparison for jet observables should contain full (or fast) 
simulation of detector responses, that has to work within the experiment's software suite and it 
certainly cannot be done within the phenomenological paper. However we try to take into account 
the basic experimental effects (affecting the jet observables) in some simplified but reasonable 
ways. In order to include a jet reconstruction on the calorimetric level in our simulation, we apply 
the iterative cluster finding algorithm PYCELL implemented in PYTHIA~\cite{Sjostrand:2006za}. Final 
state jets within cone size $R^{\rm jet}=\sqrt{\Delta \varphi^2 + \Delta \eta^2}=0.3$ with 
transverse energy $E_{\rm T}^{\rm jet}>100$ GeV and pseudorapidity $0.3<|\eta^{\rm jet}|<2$ were 
considered. The numerical results were compared with the CMS data on modification factors for jet 
shapes~\cite{Chatrchyan:2013kwa} and jet fragmentation function~\cite{Chatrchyan:2014ava} using the 
same kinematic cuts as in the experiment. Note that the data have been obtained by more complicated 
way, using the anti-$k_{\rm T}$ algorithm~\cite{Cacciari:2008gp}, utilizing so-called “particle-flow” 
objects that combine tracking and calorimetric information. However we assume that the physical 
observables should be stable for different reasonable algorithms. For example, 
in~\cite{Chatrchyan:2011sx} the calorimeter-based iterative cone algorithm was applied as a basic 
option for jet reconstruction, while the anti-k$_{\rm T}$ algorithm based on particle flow objects was 
used for a cross-check of the results. A good agreement between the results obtained with two 
algorithms has been found. 

Another simplification in our simulation is not taking into account the high multiplicity background. 
The correct performance of jet reconstruction algorithms and background subtraction procedure is
checked and verified in any such experimental analysis (including ones discussed in the current paper). 
So the measured physical observables are supposed to be stable with respect to the background fluctuations. 
Since attempt to reproduce the jet analysis procedure within the model by exactly the same way as in the experiment would require huge extra efforts without crucial influence on the results obtained, we 
believe that our simulation of experimental effects is enough to support our main physical conclusions. 

At first we have checked that the overall suppression of inclusive jet rates for 10\% of most 
central PbPb events as compared to the corresponding pp collisions (scaled to the number of 
binary nucleon-nucleon collisions) almost does not depend on $E_{\rm T}^{\rm jet}$, and 
is found on the level $R_{\rm AA}^{\rm jet} \sim 0.5 \div 0.55$ for ``Scenario {\it S}'' and 
$R_{\rm AA}^{\rm jet} \sim 0.4 \div 0.45$ for ``Scenario {\it W}''. Both results are close to  
the ATLAS measurements of the jet suppression factor~\cite{Aad:2012vca,Aad:2014bxa} within statistical 
and systematic uncertainties of the data. Since no 
qualitative difference between ``Scenario W'' and ``Scenario S'' seen for the energy dependence of 
$R_{\rm AA}^{\rm jet}$ (only numerical difference $\sim 20$\% independently of $E_{\rm T}^{\rm jet}$), making unambiguous conclusions in favour of either scenario based on $R_{\rm AA}^{\rm jet}$ measurements  
would be rather difficult. Our explanation for this is as follows. The naive expectation is that small-angle radiation will, in the first place, soften particle energy distributions inside the jet, increase the multiplicity of secondary particles, but (almost) will not affect the total jet energy. In such a case jet suppression could be significantly less pronounced than the one for wide-angle radiation. However the inevitable feature of any jet reconstruction algorithm is the separation of high and low transverse momentum particles (clusters) to treat them as extracted signal and subtracted background  respectively. Thus softening of particle distribution inside the jet effectively results in loss of jet energy, and difference between two scenarios for jet suppression factors becomes not so pronounced. 

Then we consider the jet internal structure. The radial jet profile may be characterized 
by the distribution of the transverse momentum inside the jet cone:
\begin{equation}
\rho(r) = \frac{1}{\delta r} \frac{1}{N_{\rm jet}} \sum_{\rm jets} \frac{p_{\rm T}(r-\delta
r/2, r + \delta r/2)}{E_{\rm T}^{\rm jet}}~,
\label{eq:rho}
\end{equation}
where 
$r = \sqrt{(\eta - \eta_{\rm jet})^2 + (\varphi -\varphi_{\rm jet})^2} \le R^{\rm jet}$ 
is the radial distance from jet particle to the jet axis, defined by the coordinates 
$\eta_{\rm jet}$ and $\varphi_{\rm jet}$. Following CMS analysis 
procedure~\cite{Chatrchyan:2013kwa} the jet cone was divided into six bins with radial 
width $\delta r = 0.05$, and the transverse momentum of all charged particles with 
$p_{\rm T} > 1$ GeV/$c$ in each radial bin was summed to obtain the fraction of the total 
jet $p_{\rm T}$ carried by these particles. Then the results were averaged over the total 
number of found jets, $N_{\rm jet}$.

The longitudinal jet profile usually is characterized by the jet fragmentation function 
$D(z)$ defined as the probability for a jet particle to carry a fraction $z$ of the jet transverse energy. Often jet fragmentation function is measured in 
terms of variable $\xi=\ln (1/z) = \ln (E_{\rm T}^{\rm jet}/p_{\rm T})$, 
and it normalized to the total number of jets, $N_{\rm jet}$. The charged particles with 
$p_{\rm T} > 1$ GeV/$c$ in a jet cone were selected for the 
analysis~\cite{Chatrchyan:2014ava}.   

\begin{figure*}
\begin{center}
\resizebox{1.\textwidth}{!}{%
\includegraphics{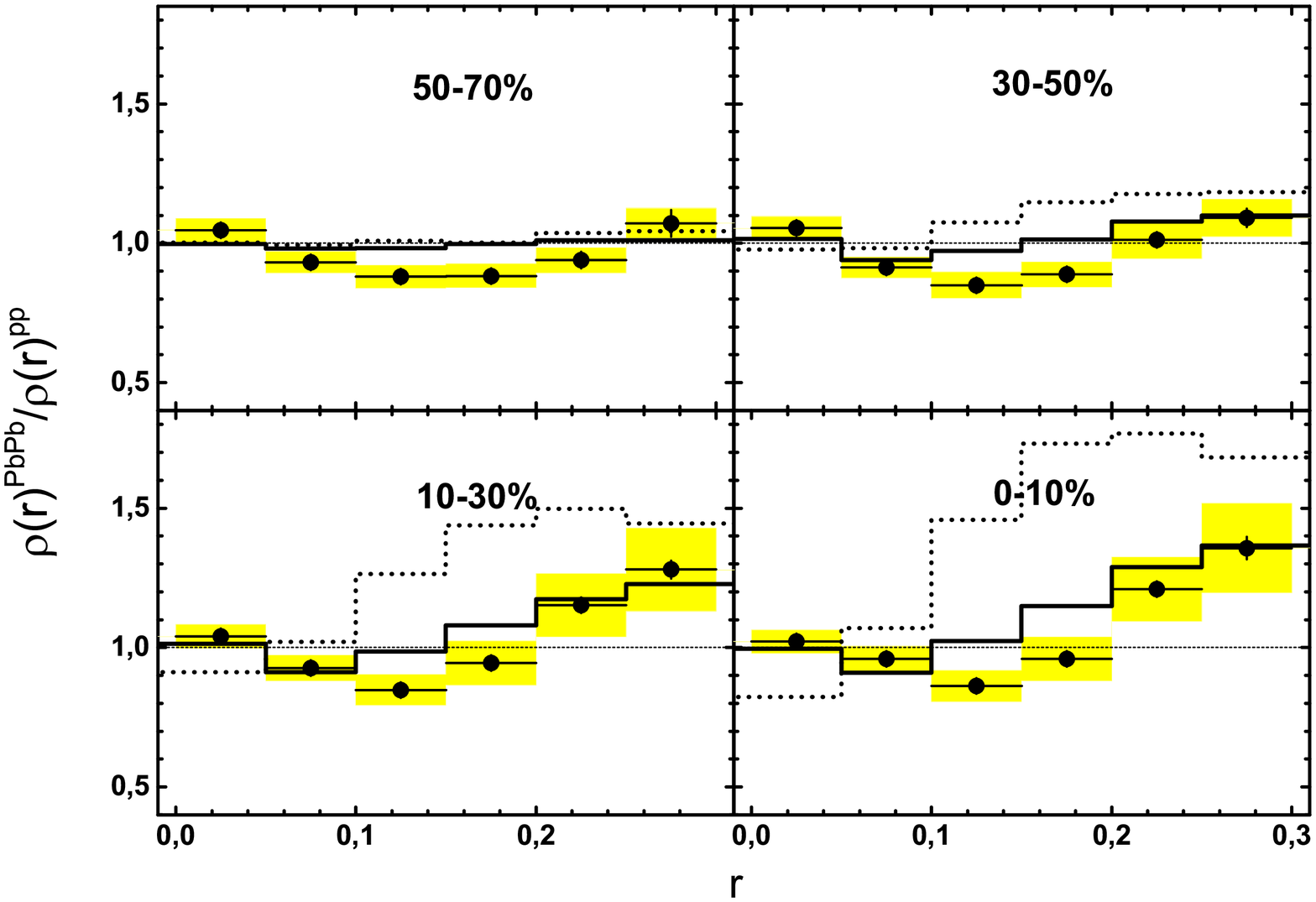}
}
\end{center}
 \caption{Jet shape nuclear modification factors as a function of distance from the jet 
axis in PbPb collisions (four centralities) at $\sqrt s_{\rm NN}=2.76$ TeV
for inclusive jets with $E_{\rm T}^{\rm jet}>100$ GeV and pseudorapidity 
$0.3<|\eta^{\rm jet}|<2$. The charged particles with $p_{\rm T} > 1$ GeV/$c$ in a jet 
cone $R=0.3$ are included. The closed circles are CMS data~\cite{Chatrchyan:2013kwa}, 
the error bars show the statistical uncertainties, and the boxes show the systematic errors. 
The solid and dashed histograms are simulated PYQUEN events for ``Scenario {\it W}'' 
(wide-angle radiative plus collisional energy loss) and ``Scenario {\it S}'' 
(small-angle radiative plus collisional energy loss) respectively.} 
\label{jet_sh}
\end{figure*}

\begin{figure*}
\begin{center}
\resizebox{1.\textwidth}{!}{%
\includegraphics{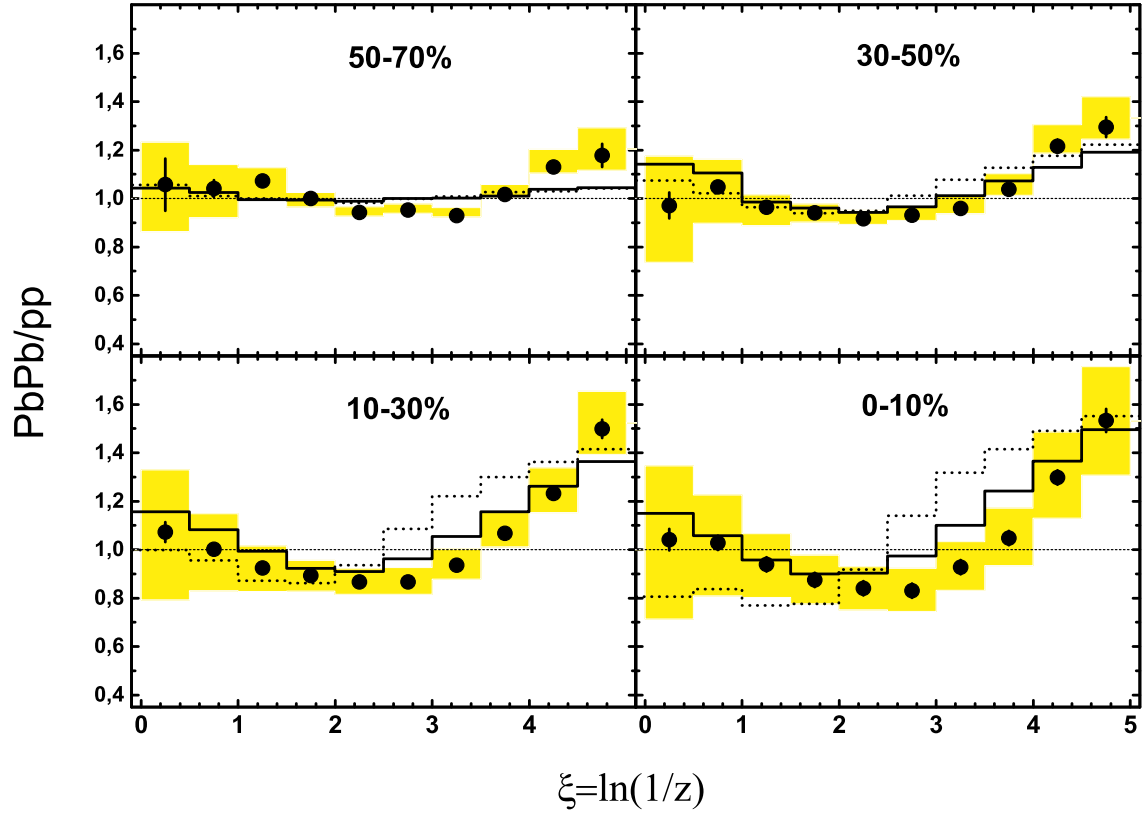}
}
\end{center}
 \caption{The ratio of jet fragmentation function in PbPb collisions (four centralities) 
to its pp reference at $\sqrt s_{\rm NN}=2.76$ TeV for inclusive jets with 
$100<E_{\rm T}^{\rm jet}<300$ GeV and pseudorapidity $0.3<|\eta^{\rm jet}|<2$. The 
charged particles with $p_{\rm T} > 1$ GeV/$c$ in a jet cone $R=0.3$ are included. 
The closed circles are CMS data~\cite{Chatrchyan:2014ava}, 
the error bars show the statistical uncertainties, and the boxes show the systematic errors. 
The solid and dashed histograms are simulated PYQUEN events for ``Scenario {\it W}'' 
(wide-angle radiative plus collisional energy loss) and ``Scenario {\it S}'' 
(small-angle radiative plus collisional energy loss) respectively.} 
\label{jet_ff}
\end{figure*}

Figure~\ref{jet_sh} shows the jet shape nuclear modification factors, 
$\rho(r)({\rm PbPb})/\rho(r)({\rm pp})$, for four centralities of PbPb collisions. 
The specific modification of radial jet profile in most central collisions due to 
a redistribution of the jet energy inside the cone is observed. It includes the excess at 
large radii, the suppression at intermediate radii, and unchanged (or slightly enhanced) 
jet core. PYQUEN (``Scenario {\it W}'') produces the similar modification close to the data 
(within the experimental uncertainties). At the same time PYQUEN (``Scenario {\it S}'') gives 
qualitatively very different result, such as excess at large and intermediate radii, and
suppression for jet core. The similar situation appears for jet fragmentation function 
(figure~\ref{jet_ff}). The same prominent features for the ratio of PbPb jet fragmentation
function to
its pp reference seen in the data and in the PYQUEN (``Scenario {\it W}''): the excess at 
low $p_{\rm T}$ (large $\xi$), the suppression at intermediate $p_{\rm T}$ ($\xi$), and the
indication on a some enhancement at high $p_{\rm T}$ (small $\xi$, the domain of leading
particles). Note that recent ATLAS data on jet fragmentation function~\cite{Aad:2014wha}
supports the presence of excess at high $p_{\rm T}$ with more confidence (but this 
comparison was done with respect to peripheral PbPb events). As with jet shapes, PYQUEN
(``Scenario {\it S}'') fails to reproduce measured jet fragmentation functions providing 
the suppression at high $p_{\rm T}$.   

\begin{figure*}
\begin{center}
\resizebox{1.\textwidth}{!}{%
\includegraphics{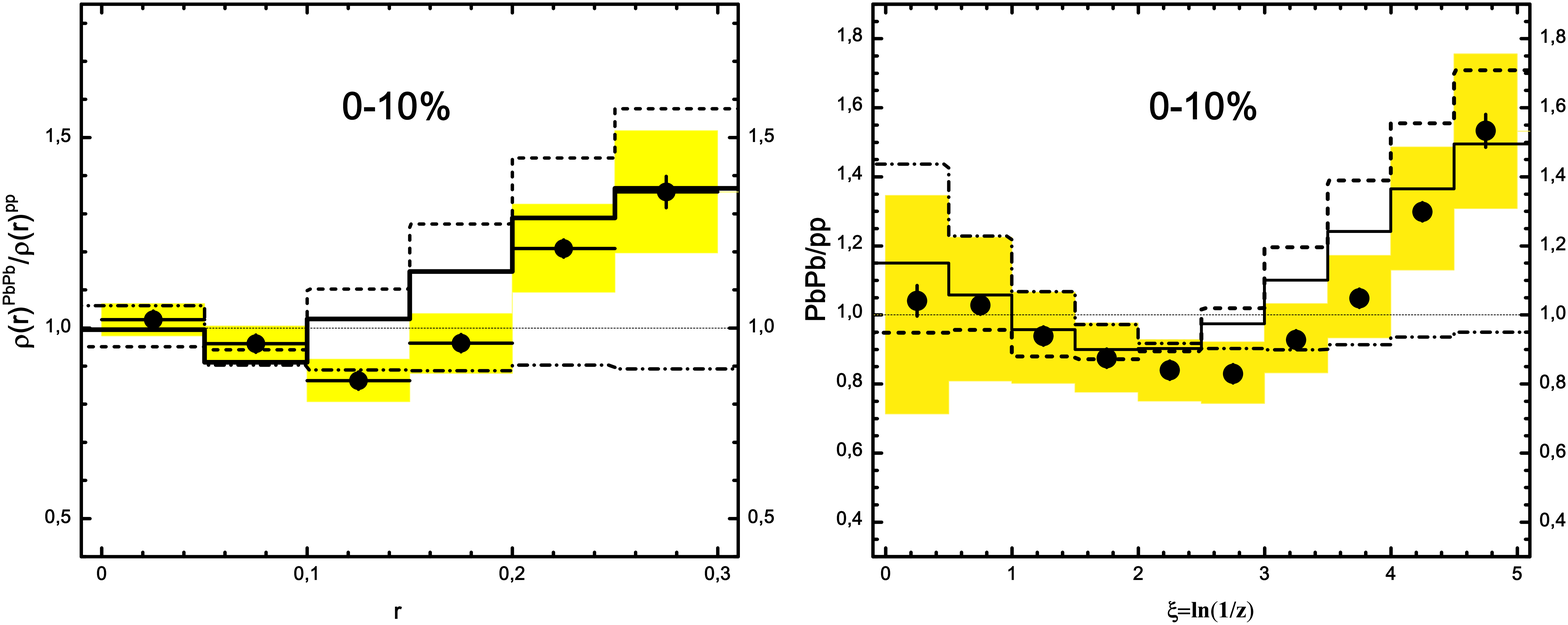}
}
\end{center}
 \caption{Jet shape nuclear modification factor as a function of distance from the jet 
axis in PbPb collisions (left) and the ratio of jet fragmentation function in PbPb 
collisions to its pp reference (right) at $\sqrt s_{\rm NN}=2.76$ TeV 
(0-10\% centrality) for inclusive jets with $E_{\rm T}^{\rm jet}>100$ GeV and 
pseudorapidity $0.3<|\eta^{\rm jet}|<2$. The charged particles with 
$p_{\rm T} > 1$ GeV/$c$ in a jet cone $R=0.3$ are included. The closed circles are CMS 
data~\cite{Chatrchyan:2013kwa,Chatrchyan:2014ava}, the error bars show the 
statistical uncertainties, and the boxes show the systematic errors. 
The solid, dashed and dash-dotted histograms are simulated PYQUEN events for 
``Scenario {\it W}'' (wide-angle radiative plus collisional energy loss), scenario   
``wide-angle radiative loss only'' and scenario ``collisional energy loss only'' 
respectively.} 
\label{rad-col}
\end{figure*}

In order to analyze the relative contribution of wide-angle radiative and collisional energy 
loss to the medium-modified intra-jet structure, two additional PYQUEN options were considered: 
wide-angle radiative loss only (without collisional loss) and collisional loss only 
(without radiative loss). The corresponding results for jet shape nuclear modification factors 
and jet fragmentation function are shown on figure~\ref{rad-col} for 10\% of most central PbPb 
collisions. As it could be expected, the contribution from wide-angle radiative energy loss 
dominates. So the option ``wide-angle radiative loss only'' provides much stronger modification 
of the jet profile than the modification obtained for the option ``collisional loss only''. 
With all this the scenario with wide-angle radiative loss alone cannot match 
the data well, so taking into account both the contributions is necessary. Note that the 
relative weight between both contributions is not regulated specially in the model, but 
implicitly is determined by the physical assumptions when calculating radiative and collisional 
energy loss (equations \ref{radiat} and \ref{col}) and by settled model parameter values 
(mostly an initial maximal temperature $T_0^{\rm max}$).

\section{Discussion}
\label{sec:discussion}

Let us discuss now the possible origin of a such specific medium-modified jet structure. 
In-medium emitted gluons are softer than initial jet partons, and fly at some 
angle with respect to the parent parton direction. Thus it is quite expectable 
that such ``additional'' gluons contribute to an excess of hadron multiplicity at low 
transverse momenta and jet radial broadening. On the other hand, the energy loss of 
initial jet partons reduces a number of hadrons at high and intermediate transverse 
momenta, such hadrons being strongly correlated with a jet axis. Then at first glance it 
should result in the suppression of hadron multiplicities in a jet core and at high 
$p_{\rm T}$, which contradicts the data. However, in fact medium-modified jet 
structure at intermediate and high $p_{\rm T}$ is determined by the interplay of two effects. 
The first one is radial broadening and longitudinal softening due rescatterings and energy 
loss of jet partons. The second one is shifting down the jet energy due 
to ``out-of-cone'' energy loss. The energy loss of a jet as a whole results in the 
difference between the ``initial'' (non-modified) and final jet energies. Since more 
energetic jets initially are more collimated and particle $p_{\rm T}$-spectrum
in such jets is more harder, two above effects enter into competition. 
If the contribution of wide-angle partonic energy loss to the total loss is 
large enough, the decrease in the yield of jet particles at high $p_{\rm T}$ and the 
broadening of a jet core can be compensated by significant jet energy 
``rescaling'', and converted into increase in the particle yield and (almost) 
unmodified radial profile at small radii. Such 
specific behavior of medium-modified jet fragmentation function at high $p_{\rm T}$ has 
been predicted some years ago in~\cite{Lokhtin:2003yq} (see also subsection 6.16.2 
in~\cite{Abreu:2007kv}). The influence of measured jet fragmentation functions by the 
enhanced quark-to-gluon jet fraction can be also important~\cite{Spousta:2013aaa}. 

Note that the parameterizations (\ref{sar}) and (\ref{war}) for gluon angular 
spectrum may be rather facile to account all features of intra-jet activity in the data, 
moreover our treatment of calorimetric jet finding being inevitably simplified as 
compared with the realistic experimental procedure. However it does not affect the main 
physical message originating from the present studies: the data support the supposition that 
the intensive wide-angle (out the jet cone) partonic energy loss occurs in most central 
PbPb collisions, while the scenario with small-angle loss seems to be inconsistent with the data.
This interpretation is also in the agreement with the results of our previous model analysis 
of medium-induced imbalance in dijet transverse energy~\cite{Lokhtin:2011qq}. As we already 
mentioned above, a specific form of angular spectrum is not so important here, only a part 
of the energy loss outside the typical jet cone affects this kind of jet observables. For 
example, in~\cite{Lokhtin:2003yq} we analyzed the relation between in-medium softening jet 
fragmentation function and suppression of the jet rates due to energy loss outside the jet 
cone without using the explicit form of angular spectrum at all. Within the current study,  
we have found that increasing the parameter $\theta_0$ in (\ref{sar}) by factor $\sim 3$ 
results in similar modification of longitudinal and radial jet profiles as it is obtained 
in ``Scenario {\it W}''. Thus since the large rise in typical radiation angle and its 
smearing in the parameterization (\ref{sar}) effectively reproduces the wide-angle radiation 
case (\ref{war}), a particular form of angular spectrum cannot be verified unambiguously 
within our simulation.  

Finally we would like refer to some other recent theoretical calculations 
for jet structure observables in PbPb collisions at the LHC. Jet fragmentation function 
was calculated and compared with the data in 
Refs.~\cite{Zapp:2012ak,Casalderrey-Solana:2014bpa,Kharzeev:2012re,Perez-Ramos:2014mna}. 
JEWEL event generator~\cite{Zapp:2012ak} and hybrid strong/weak coupling jet quenching 
model~\cite{Casalderrey-Solana:2014bpa} are successful in describing the basic trends 
seen in the data excepting low-$p_{\rm T}$ region. Hardening of the fragmentation function 
at high $p_{\rm T}$ in these models is a consequence of the depletion of softer jet 
particles. Effective 1+1 dimensional quasi-Abelian 
model \cite{Kharzeev:2012re} is able to reproduce low and intermediate $p_{\rm T}$-region, 
and predicts unmodified fragmentation at high $p_{\rm T}$. Jet fragmentation 
function~\cite{Perez-Ramos:2014mna} and jet shapes~\cite{Ramos:2014mba} were studied also with 
YaJEM event generator. It has been found that YaJEM results are qualitatively consistent with 
measurements of jet fragmentation function, while simulated jet broadening seems to be 
significantly stronger than in the data. 
 
\section{Conclusion}
\label{sec:summ}

Modification of jet fragmentation function and jet shapes in PbPb collisions at 
$\sqrt{s}_{\rm NN}=2.76$ TeV with respect to the corresponding pp data has been analyzed 
in the frameworks of PYQUEN partonic energy loss model. PYQUEN simulations with wide-angle 
radiative and collisional energy loss provide a specific medium-induced modifications of 
longitudinal and radial jet profiles in most central collisions, which are close to 
experimentally observed ones. In spite of the contribution from wide-angle radiative loss 
to the medium-modified intra-jet structure is found to dominate, taking into account 
collisional loss is also necessary to match the data. Some excess in the yield of jet 
particles at high $p_{\rm T}$ and  (almost) unmodified jet core may by explained by 
significant shifting down the jet energy due to ``out-of-cone'' energy loss. At the same 
time, the scenario with small-angle energy loss does not reproduce this effect, and so 
seems to be inconsistent with the data. We suppose that when the contribution of wide-angle 
energy loss into the total loss becomes large enough, the decrease in the yield of jet 
particles at high transverse momenta and the broadening of a jet core is compensated by 
significant jet energy ``rescaling'', and converted into increase in the particle yield 
and (almost) unmodified radial profile at small radii. Such rescaling is not possible if 
small-angular loss is the dominant source of energy loss. 

Together with other jet observables, the medium-modi\-fied jet structure seen in most central 
PbPb collisions at the LHC supports the presence of intensive wide-angle partonic 
energy loss, and can put strong constraints on the theoretical models of jet quenching. 
Future LHC data collected at increased energy and luminosity are expected to deliver more 
precise measurements of various jet characteristics in heavy ion collisions. This will allow  
us to study jet quenching mechanisms and properties of hot deconfined matter in more detail.

\section*{Acknowledgments}

\begin{acknowledgement}
Discussions with M.~Spousta and A.I.~Demianov are gratefully acknowledged.
We thank our colleagues from CMS collaboration for 
fruitful cooperation. This work was supported by Russian Scientific Fund 
(grant 14-12-00110).
\end{acknowledgement}

\begin{thebibliography}{99}
\bibitem{d'Enterria:2009am} D.~d'Enterria, Landolt-Bornstein {\bf 23}, (2010) 471
\bibitem{Wiedemann:2009sh} U.A.~Wiedemann, Landolt-Bornstein {\bf 23}, (2010) 521
\bibitem{Accardi:2009qv} A.~Accardi, F.~Arleo, W.K.~Brooks, D.~d'Enterria, V.~Muccifora, 
Riv. Nuovo Cim. {\bf 32}, (2010) 439
\bibitem{Majumder:2010qh} A.~Majumder, M.~Van~Leeuwen Prog. Part. Nucl. Phys. {\bf 66}, 
(2011) 41 
\bibitem{Dremin:2010jx} I.M.~Dremin, A.V.~Leonidov, Phys. Usp {\bf 53}, (2011) 1123  
\bibitem{Mehtar-Tani:2013pia} Y.~Mehtar-Tani, J.G.~Milano, K.~Tywoniuk, Int. J. Mod. Phys. A
{\bf 28}, (2013) 1340013
\bibitem{Majumder:2014vpa} A.~Majumder, arXiv:1405.2019 [nucl-th]
\bibitem{brahms} I.~Arsene, et al. (BRAHMS Collaboration), Nucl. Phys. A {\bf 757}, (2005) 1
\bibitem{phobos} B.B.~Back, et al. (PHOBOS Collaboration), Nucl. Phys. A {\bf 757}, (2005) 28
\bibitem{star} J.~Adams, et al. (STAR Collaboration), Nucl. Phys. A {\bf 757}, (2005) 102
\bibitem{phenix} K.~Adcox, et al. (PHENIX Collaboration), Nucl. Phys. A {\bf 757}, (2005) 184
\bibitem{Spousta:2013aaa} M.~Spousta, Mod. Phys. Lett. A {\bf 28}, (2013) 1330017
\bibitem{Aad:2010bu} G.~Aad, et al. (ATLAS Collaboration), Phys. Rev. Lett. {\bf 105}, (2010) 252303
\bibitem{Chatrchyan:2011sx} S.~Chatrchyan, et al. (CMS Collaboration), Phys. Rev. C {\bf 84}, 
(2011) 024906 
\bibitem{Chatrchyan:2012nia} S.~Chatrchyan, et al. (CMS Collaboration), Phys. Lett. B {\bf 712}, 
(2012) 176 
\bibitem{Chatrchyan:2012gt} S.~Chatrchyan, et al. (CMS Collaboration), Phys. Lett. B {\bf 718}, 
(2013) 773 
\bibitem{Aad:2012vca} G.~Aad, et al. (ATLAS Collaboration), Phys. Lett. B {\bf 719}, 
(2013) 220 
\bibitem{Abelev:2013kqa} B.~Abelev, et al. (ALICE Collaboration), JHEP {\bf 1403}, (2014) 013
\bibitem{Aad:2014bxa}  G.~Aad, et al. (ATLAS Collaboration), Phys. Rev. Lett. {\bf 114}, (2015) 072302
\bibitem{Aad:2013sla} G.~Aad, et al. (ATLAS Collaboration), Phys. Rev. Lett. {\bf 111}, (2013) 152301  
\bibitem{Chatrchyan:2013exa} S.~Chatrchyan, et al. (CMS Collaboration), Phys. Rev. Lett. 
{\bf 113}, (2014) 132301 
\bibitem{Chatrchyan:2014hqa} S.~Chatrchyan, et al. (CMS Collaboration),  Eur. Phys. J. C {\bf 74},
(2014) 2951
\bibitem{CasalderreySolana:2010eh} J.~Casalderrey-Solana, J.G.~Milhano, U.A.~Wiedemann, 
J. Phys. G {\bf 8}, (2011) 035006
\bibitem{Qin:2010mn} G.-Y.~Qin, B.~Muller, Phys. Rev. Lett. {\bf 106}, (2011) 162302 
\bibitem{Young:2011qx} C.~Young, B.~Schenke, S.~Jeon, C.~Gale, Phys. Rev. C {\bf 84}, 
(2011) 024907
\bibitem{Srivastava:2011nq} D.~Srivastava, J. Phys. G {\bf38}, (2011) 055003
\bibitem{Lokhtin:2011qq} I.P.~Lokhtin, A.V.~Belyaev, A.M.~Snigirev, Eur. Phys. J. C {\bf 71},
(2011) 1650  
\bibitem{Lokhtin:2012re} I.P.~Lokhtin, A.V.~Belyaev, L.V~Malinina, S.V.~Petrushanko, 
E.P.~Rogochaya, A.M.~Snigirev, Eur. Phys. J. C {\bf 72}, (2012) 2045 
\bibitem{Betz:2012qq} B.~Betz, M.~Gyulassy, Phys.Rev. C {\bf 86}, (2012) 024903
\bibitem{Renk:2012cx} T.~Renk, Phys. Rev. C {\bf 85}, (2012) 064908
\bibitem{Renk:2012cb} T.~Renk, Phys. Rev. C {\bf 86}, (2012) 061901
\bibitem{Renk:2013rla} T.~Renk, Phys. Rev. C {\bf 88}, (2013) 014905
\bibitem{Apolinario:2012cg} L.~Apolinario, N.~Armesto, L.~Cunqueiro, JHEP {\bf 1302}, (2013) 022
\bibitem{Zapp:2012ak} K.C.~Zapp, F.~Krauss, U.A.~Wiedemann, JHEP {\bf 1303}, (2013) 080
\bibitem{Dai:2012am} W.~Dai, I.~Vitev, B.-W.~Zhang, Phys. Rev. Lett. {\bf 110}, (2013) 142001
\bibitem{Huang:2013vaa} J.~Huang, Z.-B.~Kang, I.~Vitev, Phys. Lett. B {\bf 726}, (2013) 251
\bibitem{Zakharov:2012fp} B.G.~Zakharov, JETP Lett. {\bf 96}, (2013) 616
\bibitem{Zakharov:2013gya} B.G.~Zakharov, J.Phys. G {\bf 41}, (2014) 075008
\bibitem{Burke:2013yra} K.M.~Burke, et al., Phys. Rev. C {\bf 90}, (2014) 014909
\bibitem{Xu:2014ica}  J.~Xu, A.~Buzzatti, M.~Gyulassy, JHEP {\bf 1408}, (2014) 063
\bibitem{Mehtar-Tani:2014yea} Y.~Mehtar-Tani, K.~Tywoniuk, Phys. Lett. B {\bf 744}, (2015) 284
\bibitem{Casalderrey-Solana:2014bpa} J.~Casalderrey-Solana, D.C.~Gulhan, J.G.~Milhano, 
D.~Pablos, K.~Rajagopal, JHEP {\bf 1410}, (2014) 19
\bibitem{Chatrchyan:2013kwa} S.~Chatrchyan, et al. (CMS Collaboration), Phys. Lett. B {\bf 730}, 
(2014) 243
\bibitem{Chatrchyan:2014ava} S.~Chatrchyan, et al. (CMS Collaboration), Phys.Rev. C {\bf 90}, 
(2014) 024908
\bibitem{Aad:2014wha} G.~Aad, et al. (ATLAS Collaboration), Phys. Lett. B {\bf 739}, (2014) 320
\bibitem{Cacciari:2012mu} M.~Cacciari, P.~Quiroga-Arias, G.P.~Salam, G.~Soyez, Eur. Phys. J. C 
{\bf 73}, (2013) 2319
\bibitem{Lokhtin:2005px} I.P.~Lokhtin, A.M.~Snigirev, Eur. Phys. J. C {\bf 45}, (2006) 211
\bibitem{Sjostrand:2006za} T.~Sjostrand, S.~Mrenna, P.~Skands, JHEP {\bf 0605}, (2006) 026
\bibitem{Baier:1996kr}  R.~Baier, Yu. L.~Dokshitzer, A.H.~Mueller, S.~Peigne, D.~Schiff, 
Nucl. Phys. B {\bf 483}, (1997) 291
\bibitem{Baier:1999ds}  R.~Baier, Yu. L.~Dokshitzer, A.H.~Mueller, D.~Schiff, 
Phys. Rev. C {\bf 60}, (1999) 064902 
\bibitem{Baier:2001qw} R.~Baier, Yu. L.~Dokshitzer, A.H.~Mueller, D.~Schiff, 
Phys. Rev. C {\bf 64}, (2001) 057902  
\bibitem{Dokshitzer:2001zm} Yu.L.~Dokshitzer, D.~Kharzeev, Phys. Lett. B {\bf 519}, (2001) 199
\bibitem{Bjorken:1982tu} J.D.~Bjorken, Fermilab publication Pub-82/29-THY, 1982  
\bibitem{Braaten:1991jj} E.~Braaten, M.~Thoma, Phys. Rev. D {\bf 44}, (1991) 1298 
\bibitem{Lokhtin:2000wm} I.P.~Lokhtin, A.M.~Snigirev, Eur. Phys. J. C 
{\bf 16}, (2000) 527 
\bibitem{bjork86} J.D.~Bjorken, Phys. Rev. D {\bf 27}, (1983) 140. 
\bibitem{CasalderreySolana:2012ef} J.~Casalderrey-Solana, Y.~Mehtar-Tani, C.A.~Salgado, 
K.~Tywoniuk, Phys. Lett. B {\bf 725}, (2013) 357 
\bibitem{Ramos:2014mba} R.~Perez-Ramos, T.~Renk, Phys. Rev. D {\bf 90}, (2014) 014018 
\bibitem{Kurkela:2014tla} A.~Kurkela, U.A.~Wiedemann, Phys. Lett. B {\bf 740}, (2015) 172 
\bibitem{Blaizot:2014ula} J.P.~Blaizot, Y.~Mehtar-Tani, M.A.C.~Torres, arXiv:1407.0326 [hep-ph]  
\bibitem{Lokhtin:1998ya} I.P.~Lokhtin, A.M.~Snigirev, Phys. Lett. B {\bf 440}, (1998) 163 
\bibitem{Dokshitzer:1991wu} Yu.L.~Dokshitzer, V.A.~Khoze, A.H.~Mueller, S.I.~Troian, 
{\it Basics of perturbative QCD} (Gif-sur-Yvette, France, Ed. Frontieres, 1991)
\bibitem{Cacciari:2008gp} M.~Cacciary, G.V.~Salam, G.~Soyez, JHEP {\bf 0804}, (2008) 063
\bibitem{Lokhtin:2003yq} I.P.~Lokhtin, A.M.~Snigirev, Phys. Lett. B {\bf 567}, (2003) 39
\bibitem{Abreu:2007kv} N.~Armesto, et al., J. Phys. G {\bf 35}, (2008) 054001
\bibitem{Kharzeev:2012re} D.~Kharzeev, F.~Loshaj, Phys. Rev. D {\bf 87}, (2013) 077501
\bibitem{Perez-Ramos:2014mna} R.~Perez-Ramos, T.~Renk, arXiv:1411.1983 [hep-ph]

\end{thebibliography}
\end{document}